\documentstyle[aps,epsfig,prl]{revtex}
\def\comment#1{}
\def\cm#1{}

\everymath={\displaystyle}

\newcommand{\LL}{\Lambda}

\newcommand{\N}{N_c}
\newcommand{\sla}[1]{{\hspace{1pt}/\!\!\!\hspace{-.5pt}#1\,\,\,}\!\!}
\newcommand{\dslash}{\partial\!\!\!/}

\newcommand{\gd}[1]{\gamma_{#1}}

\newcommand{\ld}[1]{\lambda_{#1}}

\newcommand{\p}{\partial}
\newcommand{\f}[2]{\frac{#1}{#2}}
\newcommand{\tr}{{\rm tr}}
\newcommand{\be}{\begin{equation}}
\newcommand{\ee}{\end{equation}}
\newcommand{\beqn}{\begin{eqnarray}}
\newcommand{\eeqn}{\end{eqnarray}}
\newcommand{\Tr}{{\rm Tr}}

\newcommand{\s}{\sigma}

\newcommand{\SigM}{{M}}

\newcommand{\calL}{\mbox{${\cal L}$}}

\newcommand{\psibar}{\bar{\psi}}

\newcommand{\betacrit}{\beta^{\rm cr}}
\def\cm#1{}

 \def\lfrac#1#2{{{{#1}/{#2}}}}

\newcommand{\sdag}{{\scriptsize \dagger}}
\def\comment#1{}
\def\cc{{\rm c.c.}}
\newcommand{\Sbar}{~\bar{}\!\!S}

\title{Nonlinear sigma model approach for
phase disorder transitions \\ and the pseudogap phase 
in chiral Gross-Neveu,
Nambu--Jona-Lasinio models \\ and
strong-coupling superconductors }

\author{Egor Babaev\footnote {
email: egor@teorfys.uu.se  \ \ \ \ http://www.teorfys.uu.se/PEOPLE/egor/
}}

\address{
Institute for Theoretical Physics, Uppsala University
Box 803, S-75108 Uppsala, Sweden }

\begin{document}

\maketitle

\begin{abstract}
We briefly review the nonlinear sigma model
approach for the subject of increasing interest:
``two-step" phase transitions in the Gross-Neveu
and the modified Nambu--Jona-Lasinio  models at low $N$
and  condensation from pseudogap phase in strong-coupling
superconductors.
Recent success in describing 
 ``Bose-type" superconductors that possess two characterstic 
temperatures and a pseudogap above $T_c$ is the development 
approximately comparable with the BCS theory. 
One can expect that it should 
have influence on high-energy  physics, similar to impact 
of the BCS theory  on this subject. Although first generalizations of this 
concept to particle physics were made recently, these results were
not systematized.  In this review we summarize this development
and discuss similarities and differences of the appearence 
of the pseudogap phase in superconductors and the Gross-Neveu
and Nambu--Jona-Lasinio - like models. We 
discuss its possible relevance for chiral phase transition
in QCD and color superconductors.
This paper is organized in three parts: in the first
section we briefly review the
separation of temperatures
of pair formation and pair condensation  in strong -
coupling  and low carrier density superconductors
(i.e. the formation of the {\it pseudogap phase }).
 Second part is a
review of nonlinear sigma model
approach to an analogous phenomenon in the Chiral
Gross-Neveu model at small N. In the third section
we discuss the modified Nambu--Jona-Lasinio model where
the chiral phase transition is accompanied by a
formation of a phase analogous to the pseudogap phase.
%In conclusion we discuss a modified J
%of formation of a phase analogous to pseudogap phase
\end{abstract}
\section{Introduction}
Many concepts of particle physics have a close
relation to superconductivity, for example the
Nambu--Jona-Lasinio model \cite{NJLM}-\cite{ht}
was proposed in analogy to the BCS theory of
superconductivity and is considered as a low-energy effective
theory of QCD. Recently  a substantial
progress  has been made in the theory of superconductivity
in systems with strong attraction and low carrier
density. That is,
it has been observed that away from
the limits of infinitesimally weak coupling strength
or very high carrier density, BCS-like mean-field theories
are qualitatively wrong and these systems
possess along with superconductive phase
an additional phase where
there exist Cooper pairs but no symmetry is broken
due to  phase fluctuations
({\it the pseudogap phase}). What may
be regarded as an indication of the
importance of this concept to particle physics is
that  recently  the formation
of the pseudogap phase
due to dynamic quantum fluctuations at low $N$
was found in the
chiral Gross-Neveu model in $2+\epsilon $
dimensions \cite{gn1}.
%However in this
%paper we show that no direct generalization
%of result \cite{gn1} to NJL model is possible.

Separation of the  temperatures of the pair formation and
of the onset of phase coherence (pair condensation)
in strong-coupling superconductors,
 in fact, has been known already for many years (Crossover from BCS
superconductivity to  Bose-Einstein Condensation (BEC)
of tightly bound fermion pairs)
\cite{Le,N}.
Intensive theoretical study
of these phenomena in the recent years
(see for example \cite{sc}-\cite{nnnew}),
was sparked by experimental
results on
underdoped (low carrier density)
cuprates that display ``gap-like" feature
{\it above} critical temperature
$T_c$ that disappears only at a substantially higher
temperature $T^*$.
There is experimental evidence that
this phenomenon in high-$T_c$ superconductors
may be connected with precritical pairing
fluctuations above $T_c$.
At present, this crossover
has been studied by variety of
methods
and in many different models.

Because of  intimate relationship of
many problems in particle
physics to superconductivity
it seems  natural to guess that
the pseudogap may become a
fruitful concept in high energy physics too.

Below we review these phenomena in superconductors
and discuss its possible implications for QCD.

\section{ Pseudogap phase in
strong-coupling and low carrier density
theories of superconductivity}
\subsection{Perturbative results}
The BCS theory describes 
metallic superconductors
perfectly.
However, it failed to describe even qualitatively
superconductivity in underdoped High-$T_c$ compounds.
One of the most exotic properties of the latter materials
is the existence of a 
 pseudogap  in the spectrum of the normal state
well above critical temperature that from an experimental point
of view, manifests itself
as a significant suppression of low frequency spectral weight, thus
being in contrast to the exactly zero spectral weight in the case of the
superconductive gap. Moreover, spectroscopy experiments
show that a superconductive gap evolves
smoothly in magnitude and wave vector dependence to a pseudogap
in normal state. Besides that,  NMR and tunneling
 experiments indicate the
existence of incoherent Cooper pairs well above $T_c$. In principle it
is easy to guess what is hidden behind these circumstances,
 and why BCS theory is incapable of describing it.
Let us imagine for a moment that we are able
to bind electrons in Cooper pairs infinitely tightly -
obviously this implies that the characteristic temperature
of thermal pair decomposition will also be
infinitely high, but this does not imply that     the
long-range order will survive at infinitely high temperatures.
As  first observed in \cite{N},
long-range order will be destroyed in a similar way, as  say, in
superfluid ${}^4$He, i.e., tightly bound Cooper pairs,
at a certain temperature will acquire a nonzero momentum and
thus we will have gas of tightly bound
Cooper pairs but no macroscopic occupation of
the zero momentum level ${\bf q} =0$
and with it no long-range order. Thus phase diagram
of a strong-coupling superconductor has three regions:
\begin{itemize}
\item The superconductive phase where there are condensed fermion pairs.
\item The {\it pseudogap} phase where there exist
fermion pairs but there is no condensate
and thus  there is no symmetry breakdown and no superconductivity.
\item The normal phase with thermally decomposed Cooper pairs.
\end{itemize}
Of course, the existence  of bound pairs above the critical temperature
will result in deviations from Fermi-liquid behavior that
make the pseudogap phase  a very interesting object of
study.
In order to describe superconductivity in  such a system
%with a strong attractive interaction
the theory should incorporate pairs with
nonzero momentum. Thus, { \it the BCS scenario
is invalid for description of spontaneous symmetry breakdown
in a system
with strong attractive interaction or low carrier density}
(see \cite{N}, \cite{sc} and references therein). So in principle
in a  strong-coupling superconductor onset of long range order has
nothing to do with pair formation transition.
% Former process is a property
%of non-ideal Bose gas of Cooper pairs.
The existence of the paired fermions is only necessary but not sufficient condition for
symmetry breakdown.
The BCS limit is a rather exotic case
of infinitesimally weak coupling strength and high carrier density
when the disappearance of superconductivity
can  {\it approximately} be described a as pair-breaking transition.
The strong-coupling limit is another exotic
case where the temperatures of  pair decomposition and
symmetry breakdown can be arbitrarily separated. There is nothing surprising
in it: formally, in the case of Bose condensation of ${}^4He$ we can also
introduce a characteristic
temperature of thermal decomposition of the ${}^4 He$  atom;
however this does not mean that this temperature is somehow
related  to  the temperature of the Bose condensation of the
gas of $ He$ atoms. A schematic phase diagram of a superconductor
is in shown in Fig ~1.

\newpage

\vskip 5 cm
\begin{figure}[tb]
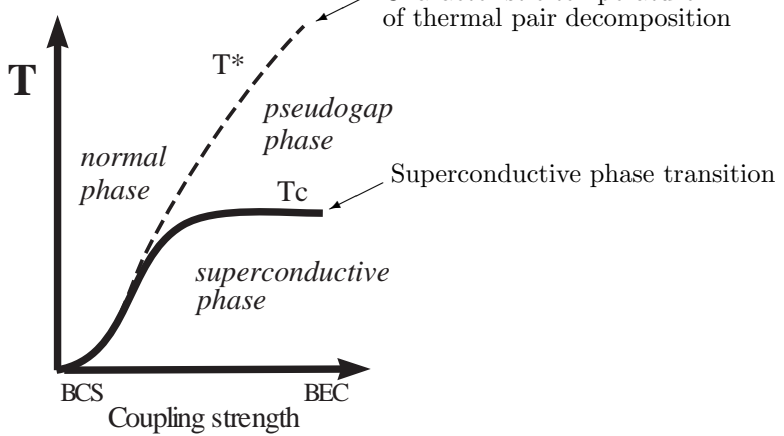

\input Phases.tps
\caption[]{Schematic phase diagram of a superconductor
with arbitrary coupling strength. In the strong-coupling limit,
the temperature of the superconductive phase transition
tends to a plateau value corresponding
to the temperature of Bose condensation of a gas of tightly
bound fermion pairs, whereas the characteristic temperature of
thermal pair decomposition grows monotonically
as a function of the coupling strength.}
\label{phases.tps}\end{figure}

Let us show how one can obtain a
pseudogap phase starting from the
BCS Hamiltonian. This was first done in the pioneering work
by  Nozieres and Schmitt-Rink \cite{N}
and in the  functional integral formalism  for 
a system with $\delta$-function attraction
by Sa de Melo, Randeria and Engelbrecht
\cite{R}. In this subsection
we briefly reproduce a part of the transparent article \cite{R}
and in the following section we
will show how qualitatively the same result can be obtained
within nonlinear sigma model (3D XY-model)
approach proposed by the author \cite{sc}.
In the following sections
 we discuss analogous nonlinear-sigma model
approach to the  similar phenomena in the chiral GN and NJL models.

The Hamiltonian of the BCS model is:
\begin{eqnarray}
H &=& \sum_\sigma \int \! d^D x
        \, \psi_\sigma^{\dag} ({\bf x})
        \left(-{{\bf \nabla}^2 \over 2m} -\mu\right)
        \psi_\sigma({\bf x})
         + g \int\!d^D x\,
        \psi_\uparrow^{\dag}({\bf x}) \psi_\downarrow^{\dag}({\bf x})
        \psi_\downarrow^{\phantom{\dag}}({\bf x})
        \psi_\uparrow^{\phantom{\dag}}({\bf x})
        \label{1.0},
\end{eqnarray}
where $\psi_\sigma({\bf x})$ is the Fermi field operator,
$\sigma=\uparrow,\downarrow$
denotes the spin components,
$m$ is the fermionic mass, and
$g < 0 $  the strength of an attractive potential
$ g  \delta ({\bf x} - {\bf x}')$.

The mean-field equations for the
gap parameter $\Delta$ and the chemical potential $\mu$
can be obtained  with a standard variation procedure:

\begin{eqnarray}
-{1\over g} &=& \frac{1}{V} \sum_{\bf k} {1\over 2 E_{\bf k}}
\tanh{E_{\bf k} \over 2T} ,\label{1.1}\\
  n &=& {1\over V} \sum_{\bf k} \left(1-{\xi_{\bf k}
 \over E_{\bf k}} \tanh{E_{\bf k} \over 2T}\right),
\label{1.2}
\end{eqnarray}
where the sum runs over all
wave vectors
$\bf k$,
 $N$ is the total number
of fermions,
 $V $  the
volume of the system,
 and
\begin{equation}
 E_{\bf k}=\sqrt{\xi_{\bf k}^2 + \Delta^2}
{}~~~\mbox{with}~~~
\xi_{\bf k} = {{\bf k}^2 \over 2 m} - \mu
\label{1.3}
\end{equation}
 are the energies of single-particle excitations.

The $\delta$-function potential produces divergence and requires
 regularization. A BCS superconductor possesses
 a natural cutoff supplied by the Debye frequency $ \omega _D$.
For the crossover problem
to be treated here
this is no longer a useful quantity, since in the strong-coupling
limit all
 fermions
participate in the interaction, not  only those
in a thin shell of width $ \omega _D$ around the Fermi surface.
To be applicable in this regime,
 we  renormalize
the gap equation in three dimensions with the help of the
%experimentally observable
$s$-wave scattering length $a_s$,
for which the  low-energy limit of the
two-body scattering process gives an equally divergent
expression \cite{R}:
\begin{equation}
        {m \over 4 \pi a_s}
=
{1\over g}
        + {1\over V}
\sum_{\bf k}
{m \over {\bf k}^2} .
 \label{1.4}
\end{equation}
Eliminating $g$ from (\ref{1.4}) and  (\ref{1.1})
we obtain a renormalized gap equation
\begin{equation}
        -{m \over 4 \pi a_s} = {1\over V} \sum_{\bf k}
        \left[{1\over 2 E_{\bf k}} \tanh{E_{\bf k} \over 2T}
 - {m \over {\bf k}^2} \right],
        \label{1.5}
\end{equation}
 in which $1/k_Fa_s$ plays the role
of a dimensionless coupling constant which monotonically increases
 from $-\infty$ to $\infty$ as the bare
coupling constant $g$ runs from small
(BCS limit) to  large values
(BEC limit).
This equation is to be solved
simultaneously with (\ref{1.2}).
These mean-field equations were
 analyzed e.g.
in Ref.~\cite{R}.

In the BCS limit, the chemical potential $\mu $
does not differ much
from the Fermi energy
 $\epsilon_F$, whereas with increasing interaction strength,
  the distribution function $n_{\bf k}$
 broadens and $\mu $ decreases, and
in the BEC limit   we have tightly bound
pairs and nondegenerate fermions with a large negative chemical
potential $|\mu|\gg T$. Analyzing  Eqns.~  (\ref{1.2}) and (\ref{1.5})
we have from (\ref{1.5})  the critical
temperature in the BCS  limit ($\mu \gg T_c$)
$T_c^{\rm BCS}=8e^{-2}e^\gamma\pi^{-1}\epsilon_F \exp(-\pi/2k_F|a_s|)$
where $\gamma=- \Gamma '(1)/ \Gamma (1) = 0.577 \dots~$,
from (\ref{1.2}) we have that chemical
potential in this case is $\mu = \epsilon_F$.
In the strong coupling limit, from
%Eq.~ (\ref{1.5}) determines $T^*$,
%whereas Eq.~ (\ref{1.2}) determines $\mu$. From
Eqs~. (\ref{1.2}),
%whereas Eq.~
(\ref{1.5})
we obtain that in the BEC limit $\mu = - E_b/2$,
 where $E_b=1/m a_s^2$ is the binding energy of the bound pairs.
In the BEC limit, the mean-field eq. (\ref{1.2})
gives that the "gap" sets in at
$T^* \simeq E_b/2 \log(E_b / \epsilon_F)^{3/2}$.
A simple ``chemical'' equilibrium estimate
$(\mu_b=2\mu_f)$ yields  for the temperature
of pair
 dissociation: $T_{\rm dissoc} \simeq E_b/\log(E_b/\epsilon_F)^{3/2}$
which shows that at strong couplings
$T^*$ is indeed related to  pair formation
\cite{R} (which in the strong-coupling regime
lies above  the temperature of the onset of
phase coherence \cite{N,R}). Obviously
$T^*$ is a monotonous function of coupling strength.
If we take into the account gaussian
fluctuations we can see  that
in the strong coupling regime, the
temperature $T^*$, obtained with the above estimate
is not related in any respect to the critical temperature
of the onset of phase coherence.
%that can
%simply by retaining Gaussian fluctuations.
Expression for the thermodynamic potential
with gaussian corrections reads \cite{N,R}:
\begin{equation}
\Omega= \Omega_0 - T \sum_{{\bf q}, i q_l }
 \ln\Gamma ( {\bf q}, i q_l),
\ee
where
\be
\Gamma^{-1} ( {\bf q}, i q_l) =
\sum_{\bf k}  \left\{  \f{1 -n_{\bf k} -n_{\bf k+q}}{i q_l-\xi_{\bf k} -
\xi_{\bf k +q}} + \f{m}{ {\bf k}^2} \right\} -\f{m}{4 \pi a_s}.
\ee
Where $n_{\bf{k}}$ is Fermi occupation and $i q_l = i T 2 \pi l $.
It is convenient following to \cite{N} to rewrite $\Omega$ in terms
of a phase shift according to definition:
 $\Gamma( {\bf q}, \omega \pm i0) = |\Gamma({\bf q}, \omega)| \exp(\pm i
\delta ({\bf q}, \omega))$. After inclusion of Gaussian correction the
number equation $N= - \partial \Omega/ \partial \mu$ reads:
\be
n=n_0(\mu,T) + \sum_{\bf q} \int_{-\infty}^\infty \f{d \omega}{\pi}
n_B(\omega)
\f{\partial \delta}{\partial \mu}({\bf q}, \omega)
\label{num}
\ee
Where $n_0$ is density of "free" fermions defined in (\ref{1.2}) and
$n_B(\omega) =1/(\exp (\omega/T)-1)$ is  the Bose function.
In order to study behavior of $T_c$ one should solve
a set of the number and gap equations.
In the BCS limit $T_c$ is not affected substantially by Gaussian
corrections, thus the  superconductive transition
can be described by mean-field theory and correspondingly $T_c \approx T^*$.
\footnote{As first discussed
in 1960s, even in  BCS superconductors there is a narrow region
of precritical pairing fluctuations.
This gives rise, e.g., to the so-called paraconductivity effect.
In particle physics, this 
phenomenon  
was pointed out
by Hatsuda and Kunihiro \cite{ht2,ht} }
In the opposite limit, numerical solution \cite{N,R} show
that the temperature of the superconductive phase transition tends
to a constant value that does not depend on the coupling strength
and is equal to the temperature of  condensation of the
ideal Bose gas of particles of mass $2m$ and density $n/2$,
where $m$ and $n$ are the mass and density of electrons correspondingly:
\be
T_c= \left[\frac{n}{2\zeta(3/2)}\right]^{2/3}\f{\pi}{m}= 0.218 \epsilon_F
\label{bose}
\ee
Where $\epsilon_F$ was used simply as a dimensional constant, namely
the Fermi energy of the gas of free fermions with the density $n$ and mass $m$
(obviously at very strong coupling
strength, when all fermions are paired there is no Fermi surface).

 The system of the gap and number equations can be solved
analytically in the strong-coupling limit. First
as it was pointed in \cite{N,R} one can make the following approximation:
to retain Gaussian corrections
only in the number equation and solve it together with mean-field
``gap" equation. Near $T_c$ in the strong -coupling regime
one finds that $\mu(T_c) = - E_b/2$, where $E_b$ is the
energy required to break a pair.
One can observe that in this limit $\Gamma({\bf q}, z)$ has an
isolated pole on the real axis for each $q$, representing
a two-body bound state with momentum $\bf q$. Since formally in this limit
we can make energy required to break a pair arbitrarily large, this
pole is widely separated from the branch cut representing
the continuum of two-particle excitations. The low energy
physics at temperatures much lower that temperature of thermal pair decomposition
is governed by this pole and one can write
$\Gamma({\bf q}, i q_m) \simeq  R ({\bf q})/[ i q_m
- \omega_b( {\bf q }) +2\mu]$,
where $\omega_b ({\bf q}) \simeq - E_b + |{\bf q}|^2/4m$.
The partition function then may be written in the following form:
\be
Z=Z_0 \int d\bar\phi d\phi \exp\left\{ \sum_{{\bf q}, iq_l }\bar\phi_q
(iq_l -\omega_b({\bf q}) + 2 \mu) \phi_q \right\}.
\ee
 Correspondingly the  strong-coupling
number equation reads:
\be
n=n_0+\sum_{\bf q}n_B[\omega_b({\bf q} ) - 2 \mu].
\ee
From which neglecting $n_0$ follows the result (\ref{bose}) \cite{N,R}
\footnote{The above estimate by \cite{N,R} when we retained
corrections only in the number equation renders correct limiting result
(\ref{bose}) however as it was observed by the authors of
\cite{N,R} in the intermediate coupling regime it gives
an artificial maximum in $T_c$ as a function of coupling strength,
this artifact is removed in higher approximations \cite{H}}.
\subsection{Nonperturbative nonlinear sigma-model
(NLSM) approach to the BCS-BEC crossover
in superconductors}
The  above discussed crossover in the BCS model
was recently studied in details perturbatively
in a  variety of approximations
(see for example \cite{H,Tch}).
Qualitatively, essential features of this crossover can be reproduced
in another simple model system - with
the help of deriving of an effective non-linear sigma
model (i.e. 3D XY-model) \cite{sc}. Moreover in the same framework of
nonlinear sigma model
one can study as well analogous crossover in 2D superconductor
\cite{Dr,Em,sh,sc}
that can not be addressed with discussed in the previous section
perturbative method
due to absence of the long-range order in 2D.
In two dimensions $T_c$ is identified with a temperature of the
Kosterltz-Thouless transition $T_{KT}$ in 2D XY-model.
In the strong-coupling (or low carrier density) regime $T_{KT}$
lies significantly lower than the temperature
of the pair formation \cite{Dr,Em,sh,sc}.
\comment{
can be relatively easily studied perturbatively in the limiting cases
of strong and weak coupling strength even though
in the above approximation one can hardly be addressed
analytically in the entire crossover region.
The reasonable question is whether there is any reason
to employ nonperturbative nonlinear sigma model
approach that is discussed below.
In fact nonlinear sigma
model approach is essentially more simple, it can
be much easier addressed analytically for
arbitrary coupling strength and carrier density, it is
appropriate for description of this phenomena
in three as well as two dimensional systems  and
as will be discussed below it provides a link
between continual and lattice models of BCS-BEC crossover.}

%In three dimensions, accuracy of nonperturbative NLSM approach can
%be verified aposteriory since we know behavior
%of critical temperature from numerical study in a variety
%of perturbative approximations (see for example \cite{H}).

As we have shown in \cite{sc}, many essential features known
from numerical study of strong-coupling
and low carrier density superconductors
are reproduced with very good accuracy
within $3D XY$-model approach.
In particular, in the framework of
NLSM approach there is  no artificial maximum
in $T_c$ in the regime of
intermedium couplings (which appears in the
approximation  discussed in the previous section).

%This circumstance
%that NLSM approach allows
%reproduce analytically all the essential propertues
%of this crossover in superconductors
%allows us to argue that a similar construiction
%in the framework of NJL model can describe a
%possible similar phenomena in QCD.
\comment{
NLSM approach provides also a link between BCS-BEC crossover
in superconductors and analogous phenomenon
in the chiral Gross-Neveu model in $2+\epsilon$ dimensions
at zero temperature that is discussed in the next section.}

%We should also note that
%in a recent paper \cite{kb} employing nonperturbative
%NLSM arguments to NJL model authors of \cite{kb}
%came to a conclusion that there is no spontaneous breakdown
%of chiral symmetry in NJL model
%due to chiral fluctuations. However
%we will show below that this result is not correct.
%Another benefit of NLSM analysis in the case of NJL
%model at finite temperature is that it provides indication
%of extreme inacuracy of mean-field derivation of
%the effective potential
%at high temperatures in the regime of low $N_c$
%and indication of
%possibile existence in NJL model a phase analogous to the
%pseudogap phase of BEC superconductors.
% and complex pair field version of Gross-Neveu model
% in 2+1 dimensions at finite temperature.
\subsubsection{XY-model approach to 3D superconductors}
Let us now reproduce results of previous subsection in the
framework of NLSM approach proposed by the author \cite{sc}.

It is very transparent  to study
the properties of the BCS-BEC crossover
to employ "modulus-phase" variables.
Following to Witten \cite{W} we can
write a partition function
as a functional integral over
modulus and phase of the Hubbard-Stratonovich field $\Delta \exp( i \varphi) $:
\begin{equation}
Z(\mu, T) = \int {\cal D} \Delta\,
{\cal D} \varphi \exp{[-\beta \Omega (T, \Delta(x), \partial \varphi
(x))]}.
\label{bk1}
\end{equation}
Assuming that phase fluctuations do not affect
local {\it modulus}
of the complex Hubbard -Stratonovich field we can
write thermodynamic potential as a sum of
``potential" and ``kinetic" (gradient)  terms:
\begin{eqnarray}
\label{ek8}
\Omega (\Delta(x), \partial \varphi(x))  \simeq
\Omega _{\rm grad} (T, \Delta, \partial \varphi(x)) +
\Omega _{\rm pot} (T, \Delta) =
\int d^3 x
%\left[
\frac{J(T, \Delta)}{2} (\nabla \varphi)^{2}
% \right],
 +
\Omega _{\rm pot} (T, \Delta).
\end{eqnarray}
Obviously in the
above expression the effective potential $ \Omega _{\rm pot} (T, \Delta) $
coincides
with the ordinary mean-field effective potential.
The gradient term
$\Omega _{\rm grad} (T, \Delta, \partial \varphi(x)) $
coincides with the Hamiltonian of 3D XY model with a stiffness $J(\Delta, T)$.

Let us reproduce low-temperature
expression for the phase stiffness in strong-coupling regime
from \cite{sc}:
\begin{equation}
J=\frac{n}{4m} - \frac{3 \sqrt{2 \pi m}}{16 \pi^2} T^{3/2}
\exp\left[-\frac{\sqrt{\mu^2+\Delta^2}}{T}
\right].
%\approx \frac{n}{4m} - \frac{3 \sqrt{2 \pi m}}{16 \pi^2} T^{3/2}
%\exp\left[- \frac{|\mu|}{T}
%\right].
\label{@stiff@}
\end{equation}
Where $n$ and $m$ are density and mass of fermions.
We can see that thermal corrections to the first
term in this regime are exponentially suppressed and the r.h.s.
tends in this limit quickly to
\begin{equation}
J_{BE}=\frac{n}{4m}.
\end{equation}
The form of this expression is not surprising - we
see that at sufficiently strong coupling strength
all fermions are bound to pairs and stiffness
becomes equal of the low-temperature 
phase stiffness of the Bose gas
of density $n/2$ and boson 
mass $2m$. Obviously
all information about internal structure of composite
bosons is evaporated from this expression in this
approximation
since at low temperature in this regime there
are no thermal pair decomposition effects
\footnote{In \cite{sc} we employed a
finite-temperature generalization of the gradient expansion at $T=0$ 
discussed in \cite{ash}.}.
%This expression also can be regarded as the
%circumstance that in BEC regime in the framework of this
%approximation we have a
%well defined 3D XY-model (i.e. influence of the
%fluctuations of the modulus of the gap function does not
%affect the stiffness coefficient in (\ref{@stiff@})).
Thus we see that
low-temperature expression
for the stiffness of the phase fluctuations reaches a plateau
value with increasing coupling strength,
whereas temperature of the
thermal pair decomposition
is a monotonously growing function of the
coupling strength.

In principle knowledge of lowest gradient term
governing gaussian fluctuations  is  not sufficient
for the study of the position of the phase decoherence transition
in a system with preformed pairs.
In continuum, 3D XY model is a free field theory and there
is no phase transition. Phase transition of 3D XY model
was studied  in great details on a lattice, so we can consider
a lattice model of BCS-BEC crossover and we can
verify aposteriory  to what extend lattice model
reproduces features of this crossover in continual model.
%This step seems of introduction of the
%lattice $3D XY$-model may seem to be somewhat cavalier
%since a lot of information of the
%systems is lost after retaining only a lowest
%gradient term and it is more properly
%to say that  below we shall speak about
%another model of the BCS-BEC crossover
%that surprisingly reproduce very accurately
%all the essential
%features of the phase diagram of the continual
%system.
Many aspects of relation of 3D  XY model
to Bose condensation, in particular derivation
of the Gross-Pitaevskii equation near $T_c$
can be found in \cite{GFCM}.

So, let us consider theory (\ref{ek8}) on a lattice
with spacing $a=1/n_{pair}^{1/3}$ where $n_{pair}$ is
concentration of Cooper pairs. This model
describes condensation of hard-core bosons on the
lattice and is a special case of Bose-Hubbard model
\footnote{In the ordinary attractive Hubbard model considered
in \cite{N} the critical  temperature is a nonmonotonous function
of the coupling strength: $T_c$ decreases in
the strong-coupling limit due to  in that model composite
bosons move via virtual ionization.}.
%Surprisingly as it will be shown below this lattice model
%reproduce nearly perfectly properties of BCS-BEC crossover
%in the initial continual theory.

Critical temperature of the phase transition of
3D XY model can be obtained with a simple mean-field
estimate \cite{GFCM} \footnote{ We should
stress that estimation  of critical
temperature of  the phase transition of the effective 3D XY model
with mean-field methods has nothing to do with BCS mean field
approximation since derivation of  the phase stiffness
coefficient (\ref{@stiff@})
required studying the gaussian fluctuations in the BCS model
\cite{sc}, and thus
this may be regarded as the approximation of the same level
as considered in the previous section.}:
\be \label{tc3d}
T_c^{3D XY} \approx  3 J a
\ee
Where $a=n_{pair}^{-1/3}$ is the lattice spacing.
%However in case if want to study position of phase decoherence
%transition  in this low temperature expression
%With it we can estimate temperature of the phase transition
%of the effective $3D XY$ model that in this model
%corresponds to the onset of phase coherence
%in a strong-coupling superconductor.
In contrast to the ordinary $3D XY$-model, in order to find
temperature of the phase transition we should study
a system of the equations (\ref{1.2}), (\ref{1.5}) , (\ref{@stiff@}) and
(\ref{tc3d}). System of these equations can be solved
however analytically
in the strong coupling limit, the result is \cite{sc}:
%]
\begin{equation}
T_c=
\frac{3}{2m} \left[ \left(\frac{n}{2}\right)^{2/3}-
\frac{1}{n^{1/3} } \frac{1}{2^{ 7/6}\pi^{3/2}} T_c^{3/2} m^{3/2}
\exp\left( -\frac{\sqrt{\mu^2+\Delta^2}}{T_c}
\right)
\right]
%\nonumber
%\\
%\simeq  \frac{3 n^{2/3}}{2^{5/3}m}-
%\frac{3}{2^{1/6} 8 \pi^{3/2} } \frac{1}{\sqrt{m}n^{1/3}}
%\exp\left( \frac{\mu}{T}
%\right).
\label{chaoslabs}
\end{equation}
With increasing
coupling strength,
this quickly tends from below to the  value (compare with (\ref{bose})):
\begin{equation}
T_c = \frac{3 n^{2/3}}{2^{5/3} m} =
 \frac{3}{(6 \pi^2)^{2/3}}\epsilon_F \approx 0.2 \epsilon_F.
\label{tcsig}
\end{equation}
Where  constant $\epsilon_F$
is Fermi energy of free fermi gas of density $n$ and fermion mass $m$.
%This approximately coincides with
%the temperature of the condensation of bosons
%of mass $2m$ and density $n/2$  (\ref{bose}).
We should observe that in nonperturbative NLSM
approach $T_c$ approaches plateau value (\ref{tcsig}),
that depends only on mass $2m$ and density $n/2$
of composite bosons,
 from below. This is in agreement
with numerical study in higher approximations
\cite{H},
whereas in the approach presented in previous
subsection $T_c$ has an artificial maximum at the intermediate
coupling strength thus approaching the limiting value from above.
Another circumstance that we discuss below is that
NLSM approach gives also qualitatively correct results in the
opposite limit of weak-coupling strength \cite{sc}.
In the weak-coupling limit near $T_c$, the stiffness coefficient
may be derived with the help of Gorkov's well-known
method :
\begin{equation}
J_{\rm BCS}=\frac{7}{48 \pi^4} \zeta(3) \frac{p_F^3}{m} \frac{\Delta^2}{T^*{}^2},
\label{@stiff}\end{equation}
This  is precisely the coefficient
of the gradient term in the Ginzburg-Landau expansion.
In the weak-coupling limit the two temperatures of
the onset of pairing correlations and the
onset of phase coherence $T^*$ and $T_c$
merge according to the formula \cite{sc}:
\begin{equation}
{T_c }= {T^* } - \frac{(2 \pi^2)^{2/3}}{2}
\frac{T^*{}^{5/2}}{\epsilon_F^{3/2}} \rightarrow T^*,
\end{equation}
%where $\alpha=(2 \pi^2)^{2/3}/2\approx 3.65$.

With it one can see \cite{sc} that in the weak-coupling
limit, the temperature of the phase transition of the
effective XY-model tends from below to
the characteristic temperature of the disappearance of the
effective potential and merges with it
for infinitesimally weak coupling strength.
Thus we arrive at some sort of aposteriori verification
of BCS behavior in this limit in the model
of hard-core composite bosons on the lattice (i.e. in this limit
if  nonzero modulus of the complex gap function
$ \Delta e^{i \varphi (x)}$
appears at some temperature, at the same
temperature  phase coherence is established and
continuous symmetry is broken).  In the weak and moderate
coupling strength limits the
disappearance of superconductivity is a competition of two
processes - pairbreaking which is thermal excitation
of individual particles and decoherence process which
is thermal excitation
of collective modes.
%Region above $T_c$
%where the pairing fluctuations are importand
%is negligible small
%in the weak-coupling superconductors and shrinks
%with decreasing coupling strength even though
%it still retains some features of pseudogap phase
%(for example paraconductivity phenomenon) \cite{asl}.
%$\Delta e^{i \varphi (x)}$
%with  nonzero gap modulus $\Delta$
%but random phase $\varphi (x)$ so the average
%$<\Delta e^{i \varphi (x)}>$ is zero
%and thus no continuous symmetry is broken,
%as it happens in the O(2) sigma-model.
%\footnote{In the case of superconductor this
%crossover also takes place as a function of
%carrier density with BCS limit corresponding to
%high carrier density}

Let us now summarize the results that follow from the NLSM
consideration. In this model in strong-coupling or
low carrier density regimes system possesses three
phases:
\begin{enumerate}
\item Superconductive phase ($T < T_c^{3D XY}$).
\item Pseudogap phase ($T_c^{3D XY} < T < T^*$) -
the phase where there exists a local gap modulus $\Delta$
that signalizes existence of the tightly bound (but noncondenced)
fermion pairs
but phase is random so average of the complex gap is
zero ($<|\Delta| \exp( i \phi)> = 0$). So in this phase
there is no superconductive gap and with it no
symmetry breakdown.
%\footnote{Term "pseudogap" is somewhat misleding
%since  even though experimentally one can observe in this phase
%an essential suppression of
%low-frequency spectral weight due to presence
%of noncondenced Cooper pairs, there is of course { \it no
%proper gap} in the spectrum.}.
\item Normal phase ($T>T^*$) the phase with
thermally decomposed Cooper pairs.
 \end{enumerate}
\subsubsection{ XY-model approach to 2D superconductors}
In two dimension there is no proper long-range order
and superconductive phase transition is associated
with a Kosterlitz-Thouless transition. In order to study
this transition it is sufficient to extract lowest gradient term
that determines temperature of the phase transition according to the formula
\cite{Dr},\cite{Em}\cite{sh},\cite{sc} \footnote{ In principle there is no
KT phase transition in a charged system due to Meissner effect, however
coupling to electromagnetic field
is always neglected in discussion of 2D superconductors, due
to experimentally in-plane penetration length in high-$T_c$
materials is much larger
than coherence length.}:
\begin{equation}
T_{\rm KT}=\frac{\pi}{2} J(\mu, T_{\rm KT}, \Delta(\mu, T_{\rm KT})).
\label{e1}
\end{equation}
This equation just like in the discussed above 3D case should
be solved self-consistently with equations for the gap modulus
and chemical potential (\ref{1.2}), (\ref{1.5}).
The result in the strong-coupling limit is \cite{sc}:
\begin{eqnarray}
T_{\rm KT} \simeq
\frac{\pi}{8} \frac{n}{m}
\left\{
1 - \frac{1}{8} \exp\left[
\frac{2\mu}{\epsilon_F} -4
\right]
\right\}.
\label{e32}
\end{eqnarray}
 Thus  for increasing coupling strength,
%i.e., decreasing
%crossover parameter $x_0 \ll -1$,
the phase-decoherence temperature $T_{\rm KT}$
tends very quickly towards a constant value \cite{Dr,Em,sh,sc}
corresponding
to KT transition in system of bosons with density $n/2$
and mass $2m$ (whereas characteristic temperature
of the thermal pair decomposition continues to grow monotonously
with the growing coupling strength):
\begin{equation}
T_{KT} =  \frac{\pi}{8} \frac{n}{m}.
\label{e302}
\end{equation}
So, this phenomenon in 2D is qualitatively similar
to the above discussed 3D case.
\vskip 1cm

There is no superconductivity in the pseudogap phase
however it
exhibits rich exotic non-Fermi-liquid behavior due to local
pairing correlations  that makes it as
interesting an object of theoretical and
experimental study as the superconductive phase itself.
In particular,
along with specific heat, optical conductivity
and tunneling experiments there are following
circumstances observed in the pseudogap phase:
In experiments on YBCO a significant suppression of
in-plane conductivity $\sigma_{ab}(\omega)$
was observed at frequencies below 500 ${\rm cm}^{-1}$  beginning
at temperatures much above $T_c$.
Experiments on underdoped samples revealed
deviations from the linear resistivity law. In particular
$\sigma_{ab}(\omega=0;T)$
increases slightly with decreasing $T$
below a certain temperature.
NMR and neutrons observations
show that below temperatures $T^*$ much higher than $T_c$,
 spin susceptibility starts decreasing.

In conclusion,
let us once more emphasis  essential features  of this phenomenon
in superconductivity:
\begin{itemize}
\item Away from a very special limit of infinitesimally
weak coupling strength and high carrier density, superconductors
are characterized by two temperatures $T_c$ and $T^* (>> T_c)$.
$T^*$ is the characteristic
temperature below which
pair correlations become important (or  in the regime
of strong interaction it is the characteristic
temperature of the formation of
real bound pairs).  $T_c$ corresponds
to the onset of phase coherence
in a system of preformed fermion pairs.  The region
of non-Fermi liquid behavior
between $T_c$ and $T^*$  calls {\it the pseudogap phase},
however the term ``pseudogap", originated in early experimental
papers, may seem somewhat misleading since, even though
a substantial depletion of low-frequency spectral weight is
observed in this region experimentally - there is
{\it no superconductive gap in the spectrum}.
\item One should note that there is { \it no
proper phase transition at $T^*$}, which
is simply a characteristic temperature of thermal
decomposition of certain fraction of noncondensed Cooper
pairs. Even though the position of this temperature
may be reasonable estimated with mean-field methods,
second-order phase transition at $T^*$ is certainly an
artifact of the discussed above approximation. Experiments  on specific
heat indicate however certain features at this characteristic
temperature.
\end{itemize}

In what follows we discuss possible implication of these
results to QCD that may posses
a phase
analogous to the pseudogap phase in strong-coupling superconductors.
The simplest model related to particle physics
that displays  pseudogap behavior of dynamical
origin is Chiral Gross-Neveu model at low $N$ that is discussed in the next
section in $2+ \epsilon$ dimensions.
\section{Pseudogap phase in Chiral Gross-Neveu model
in $2 + \epsilon$ dimensions at low N}
%%%%%%%%%%%%%
%%%%%%%%%%%%%
Let us now discuss a phenomenon similar
to pseudogap in a simple
field-theoretic model -
chiral version of the Gross-Neveu model \cite{GNM},
whose  Lagrange density is
\begin{eqnarray} \label{8.67b}
  {\cal L} = \bar\psi_a  i\sla{\partial}
       \psi _a + \frac{g_0}{2N}
\left[
\left( \bar\psi _a       \psi _a\right) ^2
+\left( \bar\psi _a  i \gamma_5     \psi _a\right) ^2
\right]  .
\end{eqnarray}
Where index $a$ runs from $1$ to $N$.
Appearance of the pseudogap phase in this model
has quite similar roots with the  above discussed
phenomenon in strong-coupling superconductors.
In superconductors we observed appearance of the
pseudogap phase on the phase diagram in the region 
 away from the limits of
infinitesimally weak coupling strength or extremely high
carrier density - i.e. in the regime when
BCS mean-field treatment is no longer valid.
Chiral Gross-Neveu model can be treated
in the limit of infinite number of field components $N$
in a framework of mean-field approach quite similar to BCS theory. Transparently in
the mean-field approximation one can find only one phase transition
at certain value of the coupling strength, similar to BCS
phase transition.  However, at  low $N$, system start
to perform dynamic chiral fluctuations which
as we have shown in \cite{gn1} give rise to  
a second, phase disorder, transition.
So at low $N$ the model possesses two
transitions at two characteristic values of
renormalized coupling constant.
% and one can drive a
%straightforward analogy to
%two characteristic temperatures $T_{KT}$
%and $T^*$ in the case of a superconductor in 2+1 dimension.
Let us reproduce this result.

One can write the collective field action for this model as:
%(\ref{8.74}) is then replaced by
\begin{eqnarray} \label{8.74b}
  {\cal A}_{\rm coll} [\sigma ] = {N} \left\{ -
           \frac{1}{2g_0} (\sigma ^2+\pi^2) - i \Tr \log
            \left[ i \sla{\partial}  - \sigma (x)-i \gamma_5\pi\right]
            \right\}.
\end{eqnarray}
This expression is invariant under the continuous set of chiral O(2)
transformations which rotate $ \sigma$ and $ \pi$ fields into each other.
This model is equivalent to
another one:
\begin{eqnarray}
            {\cal L} = \bar{\psi}_a  i \sla\partial
       \psi _a + \frac{g_0}{2N} \left( \bar \psi _a C
        \bar\psi _a^T\right) \left( \psi _b^TC\psi _b\right).
\label{8.143}\end{eqnarray}
 Here $C$ is the
matrix of charge conjugation which is defined by
\begin{eqnarray}
                 C\gamma ^\mu C^{-1} = -\gamma ^{\mu T}.
\label{8.144}\end{eqnarray}
In two dimensions, we  choose the $ \gamma$-matrices
as
  $\gamma ^0  =
 \sigma^1,
~\gamma ^1  =
 -i\sigma ^2$,
and $C=\gamma ^1.$
%
%Note that
 %
%$            \left( \bar\psi _a C \bar\psi_ a^T\right) ^\sdag
%          = \psi^T_a C \psi _a$,
%%
%implying that $g_0 <0$ corresponds to an attractive potential.
The second model goes over into the first by replacing
$
  \psi \rightarrow  \frac{1}{2}(1-\gamma _5) \psi +  \frac{1}{2}(1+\gamma _5)
          C\bar\psi ^T,
$ where superscript T denotes transposition.
In the Lagrange density (\ref{8.143}) we introduce a complex
collective field by adding
a term $
   (\lfrac{{N}}{2g_0} )\left|  \Delta - \frac{g_0}{{N}}
 \psi ^T_b C \psi _b\right| ^2,$ leading to the partition function
\begin{eqnarray}
   \!\!\!\!\!\!\!\!\!\!\!\!
\!\!\!\!\! Z[\eta,\bar\eta] = \int {\cal D} \psi {\cal D} \bar\psi
   {\cal D} \Delta {\cal D} \Delta^\sdag
        \exp\left\{ i \int d^Dx \left[ \bar\psi _a i\sla{\partial }
          \psi _a + \frac{1}{2} \left( \Delta ^\sdag \psi _a^T
          C \psi _a + \cc\right)  + \bar\psi \eta + \bar\eta
         \psi  - \frac{N}{2g_0} | \Delta |^2\right]  \right\} .
\label{8.48}\end{eqnarray}
The relation with the previous collective fields
$ \sigma$ and $\pi$ is $ \Delta= \sigma+i \pi$.
Following to BCS procedure we can fix phase of the
order parameter, then for a constant $\Delta$, the effective
action gives rise
to an effective potential that in $2+\epsilon$ dimensions
reads:
\comment{
\begin{eqnarray} \label{8.81}
&&{  \frac{1}{{N}} v(\Delta )
          = - \frac{1}{{N}} \Gamma [\Delta ]
                    = }
 \frac{1}{2g_0} \Delta^2 - \tr (1) \frac{1}{2}
       \int \frac{d^D p_E}{(2\pi )^D} \log
       \left[ p^2_E +  \Delta^2\right].
\end{eqnarray}
Performing the integral yields in $D=2+ \epsilon$ dimensions
with $ \epsilon>0$}
\begin{eqnarray} \label{8.84}
 \frac{1}{N} v(\Delta )= \frac{\mu ^\epsilon }{2 }
        \left[ \frac{\Delta^2}{g_0\mu ^\epsilon }
           - b_\epsilon \left( \frac{\Delta }{\mu }
             \right) ^{2 +  \epsilon  } \mu ^2\right],
\end{eqnarray}
where
$\mu $
is an arbitrary mass scale, and
the constant $b_\epsilon $ stands for
\begin{eqnarray} \label{8.85}
   b_\epsilon  =  \frac{2}{D} 2^{\epsilon /2} \Sbar_D
                   \Gamma (D/2) \Gamma (1 - D/2) =  \frac{2}{D} \frac{1}{(2\pi )^{D/2}}
               \Gamma (1- D/2) ,
\end{eqnarray}
which has an $ \epsilon$-expansion
   $b_\epsilon \sim -
            \left[ 1 - (\lfrac{\epsilon }{2}) \log \left( 2\pi  e^{-\gamma}
                \right) \right]/\pi \epsilon + {\cal O}(\epsilon ).$
A renormalized coupling constant $g$ may be introduced
by the equation
\begin{eqnarray} \label{8.87}
  \frac{1}{g_0 \mu ^\epsilon } - b_\epsilon \equiv \frac{1}{g},
\end{eqnarray}
so that
\begin{eqnarray} \label{8.89}
  \frac{1}{{N}} v(\Delta ) = \frac{\mu ^\epsilon  }{2  }
           \left\{ \frac{\Delta^2}{g} + b_\epsilon \Delta^2
           \left[ 1 - \left( \frac{\Delta }{\mu }\right) ^\epsilon
           \right] \right\}.
\end{eqnarray}
Extremizing this we obtain
either $ \Delta_0=0$ or
a nonzero $ \Delta_0$ that
solves the gap equation
\begin{eqnarray}
  {1} =g_0 \,\tr (1) \int
          \frac{d^Dp}{(2\pi )^2} \frac{1}{p^2+\Delta_0^2},
\label{8.161}\end{eqnarray}
in the following form:
\begin{eqnarray} \label{8.90}
  1-\frac{g^*}{g} =  \frac{D}{2} \left(
         \frac{\Delta_0 }{\mu } \right) ^\epsilon,
\end{eqnarray}
where $g^*=-1/b_ \epsilon\approx \pi \epsilon$.

In the limit $N \rightarrow \infty$ this result is exact.
On the other hand in the opposite limit of low N the system
starts to perform fluctuations around saddle-point solution
and in order to describe this system properly one should
go beyond mean-field approximation and study
propagator of the $\theta$-field - where $\theta$ is
the phase of the order parameter.

Let us consider first the case $ \epsilon=0$
where the collective field theory
consists of complex field $ \Delta$
with O($2$)-symmetry $ \Delta = |\Delta| e^{i\theta}$.
Such a system possesses  macroscopic excitations
of the form of  vortices and antivortices that
attract each other by a logarithmic
Coulomb potential.
%, just like a gas of electrons an positrons in two dimensions.
%At high values of the ,
%the vortices and antivortices
%form bound pairs.
%The grand-canonical ensemble of pairs exhibits
%quasi-long-range correlations.
%At some
 %temperature $T_c$,
%the vortex pairs break up, and the correlations
%becomes short-range.
%Such a system possesses Kostrelitz-Thouless transition.
It is known \cite{kt} that in a such
field theory involving a pure phase field $\theta(x)$,
with a Lagrange density
\begin{equation}
{\cal L}=\frac{ \beta}{2}[\partial \theta(x)]^2,
\label{@modelld}\end{equation}
where  $ \beta$ is the stiffness of the $\theta$-fluctuations,
%The important feature of the phase field $\theta$ is that it is
%a cyclic field with $\theta=\theta+2\pi$.
there is a Kostrelitz-Thouless transition when the stiffness falls below $
 \beta_{\rm KT}={2/\pi}$ \cite{kt}.

Let us return to Gross-Neveu model. Performing an expansion
around saddle point solution one can find
propagator of the $\theta$-field when $\epsilon =0$ \cite{gn1,W}:
\begin{eqnarray}
G_{\theta\theta}
 & \approx &
 \frac{i}{N}\frac{4\pi}{q^2}+{\rm regular~ terms}.
\label{8.190xx}\end{eqnarray}
%
%This implyes the following gradient term
 %in the effective action:model field theory involving of a pure phase field $\theta(x)$,
%with a Lagrange density
%
Comparing this
with the
propagator
for the model Lagrange density  (\ref{@modelld})
\begin{equation}
G_{\theta\theta}
  =
\frac{1}{ \beta} \frac{i}{q^2}
\label{@propcpomp}\end{equation}
we identify the stiffness $ \beta=N/4\pi$.

The pair version of the chiral Gross-Neveu model
has therefore a vortex-antivortex pair breaking transition
if $N$ falls below the critical value
$ N_c=8$ \cite{gn1}.

Consider now the model
in $2+ \epsilon $
dimensions
where pairs form
when renormalized coupling constant becomes
larger than the critical value $g=g^*\approx \pi \epsilon$.
In this case the expression
for the stiffness of phase fluctuations reads \cite{gn1}
\begin{equation}
 \beta
=\frac{N}{4\pi}  \left(1 -\frac{g^*}{g}\right).
\label{@stiffn}\end{equation}
What implies a KT transition in the neighborhood of two
dimensions \footnote{There is a misleading statement about 3D
case in \cite{gn1}} at:
\begin{equation}
N_c\approx 8
 \left(1 -\frac{g^*}{g}\right)^{-1},
\label{@}\end{equation}
%As $N$ is lowered below this critical value,
%the phase fluctuations of the pair field $ \Delta $ become
%incoherent and the pair condensate dissolves.
The resulting phase diagram is shown in
Fig.~\ref{ncofg.tps}.
\begin{figure}[tb]
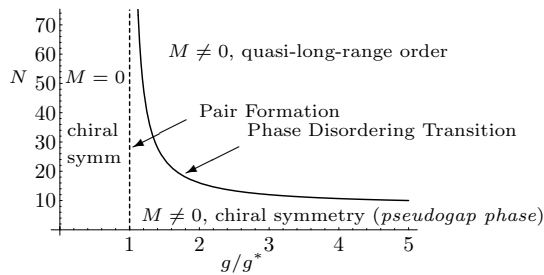

~~~~~~~~~~~~~~~
~~~~~~~~~~~~~~~
\input Ncofg.tps
\caption[]{The two transition lines in the $N-g$-plane
of the chiral Gross-Neveu model in $2+ \epsilon$ dimensions. In order
to stress difference between the local gap (i.e. ``pseudogap")
and  the order parameter analogous to the ``superconductive" gap we
denote by $M$ {\it modulus} of the order parameter ($M=|\Delta_0|$).
In this model $M$ plays the role of the ``quark" mass.
For $ \epsilon=0$, the vertical transition line coincides with the $N$-axis,
and the solid hyperbola degenerates into a horizontal line at $N_c=8$.
In the limit $N \rightarrow \infty$ generation of the quark mass
happens simultaneously with ``phase ordering" transition.}
\label{ncofg.tps}\end{figure}

In the chiral formulation of the same model in $2+\epsilon$ dimensions,
the ``pseudogap" phase has
chiral
symmetry
in spite of a nonzero spontaneously generated ``quark mass" $M=|\Delta_0| \neq 0$.
This phase is directly related to the pseudogap
phase of a strong-coupling superconductor - where
there are Cooper pairs but there is no symmetry breakdown
due to violent phase fluctuations.
The reason why this is possible is that
the ``quark mass"
depends only on $| \Delta_0|$,
thus allowing for arbitrary phase fluctuations
preserving chiral symmetry.
It is very easy to see that  the solid hyperbola in
  Fig.~\ref{ncofg.tps}
is {\it not} simply the proper (albeit approximate)
continuation of the
vertical line for smaller $N$.
There are two simple arguments.
One is formal: For infinitesimal $ \epsilon$
the first transition lies precisely at $g=g^*=\pi \epsilon$
for {\em all\/} $N$,
so that the horizontal transition line is clearly distinguished
from it (stiffness of the phase fluctuation in the regime
${g^*}/{g}\rightarrow 0$,
just like in the case of superconductor, reaches plateau
value that does not depend on the coupling
strength and $\epsilon$). The other argument is physical
and also has a clear analogy in the corresponding phenomena
in superconductivity.
If $N$ is lowered at some very large $g$, the binding energy of the
pairs {\em increases with $1/N$\/}
\{in
two dimensions,
the binding energy
is
$4M\sin^2[\pi/2(N-1)]\}$.
It is then
impossible that the
phase fluctuations on the horizontal branch of the transition line,
which are low-energy excitations,
unbind the strongly bound pairs.
%This will only happen in the limit $N\rightarrow \infty$
%where the binding energy becomes zero and the two transition
%curves merge into a single curve.
%This is the situation
%in
%the BCS theory of superconductivity, where Cooper pair binding and pair condensation
%coincide.
 Accuracy of the ``BCS" scenario in the limit $N \rightarrow
\infty$ is clearly seen from the form of phase stiffness which
has a factor $N$ that ``freezes" the phase fluctuations in this limit
and thus all the physics is essentially governed by the size of the
gap modulus.

The $2+1$-dimensional Chiral Gross-Neveu model \cite{park}
also exhibits an analogous  behavior at finite temperature \cite{gn2} 
where a similar effect is governed by thermal fluctuations. 
At finite $N$ the temperature of KT transition deviates from  
 mean-field temperature of the gap modulus formation,
 however in D=2+1 the phase diagram is substantially 
different from the phase diagram of the same model 
in $D=2+\epsilon$ at $T=0$ (see detailed 
discussion in \cite{gn2})\footnote{It should be emphasized that the
existence of the
pseudogap phase in 2+1 dimensions due to thermal
fluctuations in the Chiral Gross-Neveu model can not be rigorously proven in contrast
to $2+\epsilon$-dimensional case discussed in this section
(when there are two small parameters $\epsilon$ and $1/N$).
The 2+1 dimensional problem lacks a small parameter 
which would allow to estimate  accurately the 
position of $T^*$ at very small $N$,
however it can be argued that rough low-$N$ estimates 
show the  appearance of thermal-fluctuations-induced pseudogap phase
in this model, especially pronounced at $N\leq4$. }.

\section{Chiral fluctuations in the the NJL model at zero temperature}
Recently an attempt   was made
\cite{kb} to generalize to the NJL model
the nonlinear-sigma approach  for description of chiral
fluctuations proposed in \cite{gn1,sc}.
The authors \cite{kb}
claimed that at $N_c=3$ the NJL
model does not display spontaneous  symmetry breakdown
due to chiral fluctuations.
We show below that the
NLSM approach does not allow
 to prove that  chiral symmetry is
always restored
by  fluctuations in the NJL model at $N_c=3$.
Below we also discuss differences  from the chiral GN model,
where  the NLSM approach
allows one to reach a similar conclusion at low $N$.

The Lagrangian of the NJL model reads \cite{NJLM}
\be
\calL=\psibar
i\dslash
\psi+\f{g_0}{2N_c}\left[
\left(
\psibar\psi
\right)^2+\left(
\psibar\ld{a}i\gd{5}\psi
\right)^2
\right].
\label{NJLModel}
\ee
The three $2\times2$-dimensional
matrices $ \ld{a}/2$, generate the fundamental representation
of flavor $SU(2)$, and are normalized by
$\tr (\ld{a}\ld{b})=2\delta_{ab}$.
One can introduce Hubbard - Stratonovich fields
$ \s$ and $\pi_a$:
\be
\calL=\psibar\left(
i\dslash-\s-i\gd{5}\ld{a}\pi_a
\right)
\psi-\f{\N}{2g_0}\left(
\s^2+\pi_a^2
\right).
\label{hsnjl}
\ee
After integrating out quark fields, following 
a standard mean-field variation procedure
% standard mean-field variation
%procedure gives the gap equation:
%\be
%\tr_{\gamma}\tr_f \left[ G(x,x) {1 \choose i\ld{a}\gd{5}} \right]
%= \f{1}{g_0}{\s(x) \choose \pi_a(x)}.
%\label{gapp}
%\ee
one can choose the pseudoscalar solution $\pi_a$ to vanish
and the scalar solution $\sigma\equiv M$ to be given by a gap equation:
% analogous to
%the gap equation in a BCS superconductor:
\be
\f{1}{g_0}=i(\tr_f 1)(\tr_{\gamma} 1)\int
\f{d^Dp}{(2\pi)^D}\f{1}{p^2-M^2}
\label{gap1}
\ee
The momentum integral is regularized by means of a cutoff $\Lambda$.
\comment{
Mean-field  $N_c\rightarrow \infty$
treatment gives an effective action
\be
 \Gamma (\rho)=- \Omega [\Delta v( \rho )+v_0]
\label{EffectivePotential}
\ee
where $\Omega$ is the spacetime volume, and $v_0$
is energy density of the symmetric state,
whereas
\beqn
 &&\Delta v(\rho)=\f{N_c}{2}\Bigg\{
\f{1}{g_0}\rho^2-\f{2}{(2\pi)^2}\Bigg[
\f{\rho^2\Lambda^2}{2}
+\f{\Lambda^4}{2}\ln\left(
1+\f{\rho^2}{\Lambda^2}
\right)\nonumber\\
&&\mbox{}-\f{\rho^4}{2}\ln\left(
1+\f{\Lambda^2}{\rho^2}
\right)
\Bigg]
\Bigg\}
\label{lambdapotential}
\eeqn
the mean-field condensation energy at {\it constant}
$  \sigma ^2+\pi_a^2\equiv  \rho ^2$.
The momentum integral is regularized by means of a cutoff
$\Lambda$.
The condensation energy is extremal
at
$ \rho = \SigM$ which solves the {\em gap equation\/}
\beqn
 \f{1}{g_0}&=&\f{2}{(2\pi)^2}
\left[
\Lambda^2- \SigM^2\ln\left(
1+\f{\Lambda^2}{\SigM^2}
\right)
\right].
\label{lambdagap}
\eeqn    }
The constituent quark mass $\SigM$
in the limit $\N\rightarrow \infty$
is analogous to the superconductive gap
in the BCS limit of the theory of superconductivity.

At finite $N_c$ one can study fluctuations
around the saddle point solution.
The quadratic terms of expansion
around  the saddle point are:
\begin{eqnarray}
{\cal A}_0[\s',\pi'] = \f{1}{2}\!\int\!\!d^4q\!\left[
\left(
%\begin{array}{c}
{ \pi'_a(q) \atop
\s'(q)}
%\end{array}
\right)^T
\left(
%\begin{array}{cc}
{G_{\pi}^{-1} \ \ \ 0 \atop \
 \ 0 \  \ \ \ \ G_{\s}^{-1}}
%\end{array}
\right)\left(
%\begin{array}{c}
{\pi'_a(-q)\atop
\s'(-q)}
%\end{array}
\right)
\right],
\label{ao}\end{eqnarray}
where
$(\s',\pi'_a)\equiv(\s- \SigM,\pi_a)$
and $G_{\s,\pi}^{-1}$
are the inverse bosonic propagators.
\comment{
\begin{equation}
G_{\s}^{-1} =
\N\left[ 2\times2^{D/2} \int
\f{d^4p_E}{(2\pi)^4}\f{(p_E^2+p_Eq_E - M^2)}
{(p_E^2 + M^2)[(p_E+q_E)^2 +M^2]}
- \f{1}{g_0}\right];
G_{\pi}^{-1} =
\N\left[ 2\times2^{D/2} \int
\f{d^4p_E}{(2\pi)^4}\f{(p_E^2+p_Eq_E + M^2)}
{(p_E^2 + M^2)[(p_E+q_E)^2 +M^2]}
- \f{1}{g_0}\right].
%\label{@gre1}
\end{equation}
In the above expression one should introduce
a momentum cutoff $\Lambda_2$.} Implementing  a momentum
cutoff $\Lambda$, we can write
$G_{\pi,\sigma}^{-1}$ for small $q_E$ as:
\beqn
\!\!\!\!\!\!G_{\pi}^{-1}\!\approx\!\!-\f{\N}{(2\pi)^2}
\!\left[
\ln\left(
1\!+\!\f{\Lambda^2}{\SigM^2}
\right)
\!-\!\f{\Lambda^2}{\Lambda^2\!+\!\SigM^2}\right]
\!q_E^2\!\equiv\!\!-Z(\SigM/\Lambda)q_E^2; \ \ \ \ \ \ \ \ \
 G_{ \sigma }^{-1}\!\approx\!\!
-Z(\SigM/\Lambda)(q_E^2+4\SigM^2).
\label{SigPropStiffLambda}
\eeqn
In analogy to $3D \ XY$-model approach to
strong-coupling superconductivity \cite{sc}
the authors of \cite{kb} introduced
a unit vector field
$n_i\equiv (n_0,n_a)\equiv(\s,\pi_a)/ \rho $
and set up an effective nonlinear sigma-model
\be
{\cal A}_0[n_i]= \f{\beta}{2}\int d^4x
[\partial n_i(x)]^2.
\label{@prop}
\ee
The prefactor $\beta=M^2 Z(M/\Lambda)$,
 that follows from Eqs. (\ref{ao}) and
(\ref{SigPropStiffLambda})
plays the role of the stiffness
of the unit field fluctuations.
%
%\be
%\label{StiffModel}
%\ee
%
%

Now let us observe
that from the arguments given in
\cite{kb} it does not follow that the NJL model necessarily
remains in a chirally symmetric phase
at $N_c=3$.
At first, in contrast to the
 ($2+\epsilon$)-dimensional case, discussed in \cite{gn1},
one can not make unfortunately, any similar calculations in a
closed form in $3+1$-dimensions because this
theory is not renormalizable.
%And stiffness coefficient
%should depend on two parameters, say $M$ and $\Lambda$.
%Whereas stiffness coefficient in renormalizable theory such
%as superconductor can be expressed via one parameter.
It was already observed in \cite{cut}-\cite{bub} that
the cutoff of meson loops cannot be set equal to the  cutoff
for quark loops and thus the $1/N_c$ corrected theory
\cite{bub} possesses two
independent  parameters that may be adjusted at will.
We present another argument
of a different nature
rooted in the nonuniversality of the
critical stiffness of a NLSM in four dimensions,
%codel in $3+1$-dimensions
%of more general nature,
which does not allow one
to reach the conclusion of \cite{kb}
in the  framework
of the NLSM approach.
Our observation
also applies to the NLSM
description of precritical fluctuations in general systems.
It also allows us to show that the  additional
cutoff discussed below can not be related to the inverse coherence length
of the radial fluctuations in the effective potential as  suggested in
\cite{prl,kb}.

The  authors of \cite{kb} by deriving
$G_{\pi, \sigma}$ have essentially extracted two
characteristics from the initial system:
the stiffness of the phase fluctuations in the degenerate minimum
of the effective potential
and the mass of the radial fluctuation. However, knowledge of these
characteristics does not allow one in principle to judge
if directional fluctuations will destroy
long range order or the system will possess  a BCS-like
phase transition.
The reason is that the critical stiffness of the nonlinear sigma
model is not an universal quantity in $3+1$-dimensions.
% and
%is expressed via an additional cutoff $\Lambda_3$
%in the gap  equation (\ref{CriticalStiff}).
So in principle knowledge of the stiffness of NJL model
is not sufficient for finding the
{\it position} of the phase transition in the effective
nonlinear sigma model.
The situation is just like that in a Heisenberg
magnet, where the critical temperature depends
on the stiffness along with lattice spacing and lattice
structure. Thus if one is given only a
stiffness coefficient one can not determine the 
temperature of the phase transition\comment{Due to this
reason one can not refer to lattice simulation for
finding the value of the critical stiffiness as it was done in \cite{kb,prl}
since implicitly these numerical values contain information
of lattice structure and are not universal.}.
The situation is
in contrast to the 2D case where the  position of a
KT-transition can be deduced from the stiffness coefficient \cite{kt}.
\comment{
In two dimensions the critical stiffness of the O(2) nonlinear
sigma model is a universal quantity and is given by
$\beta_{KT}=2/\pi$ \cite{kt}, so by comparing
it with the stiffness coefficient
derived from the initial theory
(the phase stiffness of the chiral GN model in $D=2$ is
$\beta = N/4\pi$ ),
one can judge if the
system has enough phase stiffness  to
preserve quasi-long range order
as we have shown in \cite{gn1}
%\footnote{There is a misleading
%statement about three dimensional case in \cite{gn1}.}. 
That is, one can determine
the number of field components N
%and renormalized
%couling strength
 that is needed to remain below the
position of  the Kosterlitz-Thouless transition.
This is
in contrast to the  $D=3+1$ case.}
%where given stiffness coefficient one can not make
%such evaluations in general { \it in principle}.
%The authors of \cite{kb,prl}
%argued that on can
%introduce in the $3+1  \ D$ theory an
%universal critical stiffness by relating the stiffness
%to the radial.
% Obviously
%such a modified theory
%has nothing to do with initial NJL model
%where critical stiffness is not universal.
%However we show below that these
%arguments are not consistent and do
%not lead to a modified theory with an
%universal critical stiffness either.

Let us  recall a procedure for
expressing the critical stiffness
of the O(4)-nonlinear sigma model via
an additional parameter:
one can relax the constraint $n_i^2 =1$
and introduce an extra integration over the
lagrange multiplier $\lambda$, rewriting Eq. (\ref{@prop}) as:
%
%\be
$(\beta/2) \int d^4x
\left\{ [\partial n_i(x)]^2+ \lambda \left[ n_i^2(x)-1\right] \right\}$.
%\ee
%
Integrating out the $n_i(x)$-fields, yields:
\be
{\cal A}_0[\lambda]=-\beta\int d^4x  \f{\lambda(x)}{2}+\f{N_n}{2}\Tr\ln\left[
-\p^2+\lambda(x)
\right],
\label{newaction}
\ee
where $N_n$ is the number of components of $n_i(x)$ and $\Tr$ denotes the
functional  trace.
This yields a gap equation:
\be
\beta=N_n\int \f{d^4k}{(2\pi)^4}\f{1}{k^2+\lambda} .
\label{@secge}\ee
The model has  a phase transition at a  critical stiffness
that depends on an unspecified  additional cutoff parameter that
should be applied to the gap equation:
\be
\betacrit=N_n\int \f{d^4k}{(2\pi)^4}\f{1}{k^2}.
\label{CriticalStiff}
\ee
For example, in the case of magnets the additional cutoff
needed in Eq. (\ref{CriticalStiff})
is naturally related to the lattice spacing.
%The authors of \cite{kb} proposed that
%one can fix in the above equation a cutoff
%according to physical considerations.
%derived from NJL model
%that one can
%use in the above equation a cutoff
%derived from NJL model
In \cite{prl} a
criterion was proposed that states that one can relate the inverse
coherence length extracted from  radial fluctuations in an
effective potential of an initial theory to the cutoff
in the integral (\ref{CriticalStiff}) so that all the parameters
in a theory would be expressed from quantities
derived from an initial model, and thus this
modified model  possesses a universal
critical stiffness.
However, unfortunately there is no reason
for relating the cutoff needed in Eq. (\ref{CriticalStiff})
to the coherence length of the modulus fluctuations
 and moreover we show that
this procedure leads in general to unphysical consequences.
It should also be observed that it 
does not make the theory  consistent 
anyway because of  the following
circumstance:
The authors of \cite{kb} suggested  that, ``since pions in the symmetry
broken
phase are composite they are not ``defined" over the
length scales much shorter than the inverse
binding energy of the pair wave function which is
equal to $2 M$". Thus the authors of \cite{kb} performed integral
in (\ref{CriticalStiff}) up to cutoff $4M^2$.
However, since,
%it does not make the theory consistent since
as suggested in \cite{kb},
the pion fields are not ``defined" over the
length scales much {\it shorter} than the inverse
binding energy this may only serve as an
estimate for ``{\it upper} boundary" of what would
be the universal critical stiffness value.
So, unfortunately
 one can not make the conclusion of absence of symmetry breakdown
in such a modified theory by observing that
stiffness derived from the initial model is
smaller than the  maximal
possible value of would be universal critical stiffness.
%Obviously there is no reason for
%concluding from it that NJL model
%does not display symmetry breakdown
%and such a modified theory
%has nothing to do with initial NJL model.
%Moreover
%we illustrate below that such a procedure
It was also supposed in \cite{prl}
that the relation of the coherence length to the cutoff
in the equation (\ref{CriticalStiff})
yields a universal criterion
for judging the
nature of symmetry breakdown in
general physical systems.
%leads to a number of unphysical consequences.
There is a simple counterexample:
in the case of  a
strong-coupling superconductor, the
effective nonlinear sigma model
that describes
fluctuations in a degenerate valley
of the effective potential is a 3D XY-model.  In
the continuous case it is  a free field
theory and has no phase transition at all.
The phase transition appears only in the lattice theory
and, of course its temperature
depends on the lattice spacing.
As we discussed above with increasing coupling strength the
low-temperature phase stiffness of the effective 3D XY model tends
to a plateau value
$J=n/4m$, where $n$ and $m$ are the density and the mass
of fermions \cite{sc}. Thus the temperature of the phase transition
of the effective 3D XY-model is
\be
T_c^{3D XY} \propto  \frac{n}{m} a,
\label{xy}
\ee
 where
$a$ is the lattice spacing.
To be careful one should remark that accurate analysis shows
that
a strong coupling superconductor possesses two characteristic length
scales:  size of the Cooper pairs  which tends to zero with increasing
coupling strength,
and a coherence length that tends to infinity with increasing coupling
strength as the system evolves towards a weakly nonideal gas
of true composite bosons \cite{R,pist}.
First if one relates the constant $a$ in (\ref{xy}) to the size of
the Cooper pairs
%n the strong coupling limit
%ize of the Cooper pairs becomes smaller
%ith increasing coupling strength
%\footnote{
%Let us observe that in BCS superconductors
%the Cooper pair size is determined by the
%procedure equivalent to
%used in \cite{kb,prl} i.e. from the mass of the radial
%fluctuations in the effective potential}
following to the arguments of \cite{kb}
one will come to an incorrect conclusion of the absence of 
superconductivity in strong-coupling superconductors,
in a similar way as the authors
of \cite{kb} came to the conclusion
of the nonexistence of symmetry
breakdown
in  the NJL model.
This is in a direct
contradiction with
behavior of the strong coupling superconductors discussed above.
Second, if one attempts to relate $a$ in (\ref{xy})
to the second length scale
of the theory,  namely, the true coherence length, which
tends to infinity with increasing coupling strength,
then one will also come to a qualitatively incorrect conclusion 
\cite{rem}. 
%Thus the existence of an universal NLSM-based
%fluctuations criterion \cite{prl} appears to be incorrect.

Thus, in general, the nonlinear sigma model approach
for precritical fluctuations possesses an additional
fitting parameter, which is the
cutoff in the gap equation (\ref{CriticalStiff}),
which can not be related to the inverse coherence length
extracted
from radial fluctuations in an effective potential.
Thus within the NLSM approach  one can
not prove if the NJL model displays
necessarily the directional fluctuations-driven
 restoration of  chiral
symmetry at low $N_c$.
%Let us for a moment do not take
%into account above "no-go"
%arguments and observe one more
%internal controversy
%in the discussion \cite{kb}, namely the claim
%that "this approach invalidates studies of
%temperature dependence of symmetry broken phase" .
%Let us follow the lines of \cite{kb} and observe
%how the system will behave at finite temperatures
%in the framework of the approach \cite{kb}.
\section{Chiral fluctuations at finite temperature and a modified NJL
model with a pseudogap}
This section is based on the paper \cite{mprd}.
The authors of \cite{kb} employed NLSM
arguments in an attempt to show that the NJL model cannot serve
 the study of the chiral symmetry breakdown. We have shown in the
above that this conclusion appears to be incorrect
since the critical stiffness
in 3+1-dimensions is not a universal quantity and one
has an additional fitting parameter.
% that should be chosen
%from phenomenological considerations.
This is an inherent feature of the discussed NLSM approach
 in 3+1 dimensions
% and does not depend
%on whether the theory is renormalizable or not
(compare with the 
cutoffs
discussions in nonrenormalizable models in a different approach
\cite{cut}-\cite{bub}, and also \cite{bl}).
The above circumstance allows one to fix the critical
stiffness from phenomenological considerations.
However,  we argue below that, what is missed
in \cite{kb} is that, in principle, the low-$N_c$
fluctuation instabilities, when properly treated,
have a clear physical meaning.
Moreover, we argue that
 one can employ a NLSM
for describing  the chiral fluctuations 
(e.g. at finite temperature),
provided that special care is taken of
the additional cutoff parameter.
Indeed, it was already discussed in the literature that at finite temperatures
the chiral phase transition should be accompanied by
developed  fluctuations (\cite{ht,ht2} and references therein).
We argue that this
process at low $N_c$ should give  rise to a
phase analogous to the pseudogap phase that may be conveniently
described within a nonlinear sigma model approach. There are
indeed other ways to describe these phenomena,
however the NLSM
approach  seems to be especially convenient. The description of
the two-step chiral phase transition and appearance of the
intermediate phase requires one to study the system at the next-to-mean-field
level. Unfortunately, the NJL model is not renormalizable
and does not allow  one to make any conclusions about the
importance of fluctuations in a closed form \cite{bub}.
On the other hand, a pseudogap phase
is a general feature of Fermi systems with composite bosons.
The NLSM construction discussed below, because of 
its nonperturbative nature, can not be
regarded as a regular approximation but may be
considered as a tractable modification of the NJL model that
has a pseudogap.
One can also find an additional
motivation for  employing these arguments in the fact that
the NLSM approach allows one
to prove the existence of the phase analogous to pseudogap phase
in the chiral GN model \cite{gn1,gn2}
 which is the closest relative of the NJL model.
%but allows to show appearance of a pseudogap rigorously.
Also the NLSM approach works well for the description of
precritical fluctuations in superconductors \cite{sc} - where
essentially the same results have been obtained 
with different methods and in different models.
We stress that
these phenomena are a general feature
of any Fermi system with attraction.
Also, to a certain extent similar crossovers
are known in a large variety of condensed matter systems.
In particular, besides  superconductors
we might mention the
exitonic condensate in
semiconductors, 
Josephson junction arrays,
 itinerant and local-momentum theories
of magnetism and ferroelectrics.

Let us now consider the chiral fluctuations in the NJL model
at finite temperature.
Then,
% in the presence of heat resevoir
 following standard dimensional reduction
arguments [see e.g. \cite{Wil}], the chiral fluctuations should be
 described by a  $3D \ O(4)$-sigma model.
Thus one  has
the following gap equation for the effective NLSM
(i.e. the finite temperature analog of (\ref{@secge})):
\be
\frac{J_T}{T} = N_n \int \frac{d^3 k}{(2\pi)^3} \frac{1}{k^2+\lambda}
\label{gt}
\ee
The temperature of the phase transition of the three dimensional
classical $O(4)$ sigma
model with stiffness $J_T$ is expressed via the additional parameter
${\tilde \Lambda}_T$ needed in (\ref{gt}) as :
\be
T_c =  \frac{\pi^2}{2}\frac{J_T}{{\tilde \Lambda}_T}
\label{tc}
\ee
The stiffness of thermal fluctuations
$J_T$ can be readily extracted from the NJL model.
At finite temperature the inverse bosonic propagator of the collective
field $\pi$ for small $q$ can be written as:
\begin{eqnarray}
G^{-1}_\pi &= & -2^{D/2} N_c \int \frac{d^3p}{(2\pi)^3} \sum_n
\left[ \frac{T}{(p^2+M^2+\omega_n^2)^2}\right] q^2 =
%\ \ \ \ \ \ \ \ \ \
%\ \ \ \ \ \ \ \ \ \ \ \ \ \ \ \ \ \ \ \ \ \ \ \ \ \ \ \ \
%\ \ \ \ \ \ \ \ \ \ \ \ \ \ \ \ \  \  \ \ \ \ \ \ \ \ \ \ \
\nonumber \\
& - &  2^{D/2} N_c \int \frac{d^3 p}{ (2 \pi)^3}
\left[ \frac{1}{8}
\frac{1}{(p^2+M^2)^{3/2}}
\tanh
\left( \frac{\sqrt{p^2+M^2}}{2 T}\right)
-\frac{1}{16 T}\frac{1}{p^2+M^2} \cosh^{-2}
\left(
\frac{\sqrt{p^2+M^2}}{2T}
\right)
\right] q^2 = \nonumber \\
& - &  K (T,\Lambda_T, M, N_c) q^2,
\label{st0}
\end{eqnarray}
where $\Lambda_T$ is a momentum cutoff.
The propagator (\ref{st0}) gives the
gradient term that allows one to set up an effective
classical $3D ~ O(4)$-nonlinear sigma model:
\be
E=\frac{J_T (T, \Lambda_T, M, N_c)}{2} \int d^3 x [\partial n_i (x)]^2,
\label{Hei}
\ee
where
\be
J_T(T, \LL_T, M, N_c) = K(T, \LL_T, M, N_c) ~M^2(T,\Lambda_T)
\label{st1}
\ee
is the stiffness of the thermal  fluctuations in the degenerate
valley of the effective potential. The temperature -dependent quark
mass $M$ that enters this expression
is given by a standard mean-field gap equation which
also should be regularized with the cutoff $\Lambda_T$:
\be
\f{1}{g_0}=2\times2^{D/2}\sum_n
\int \f{d^3p}{(2\pi)^3}\f{T}{p^2+ M^2 +\omega_n^2}.
\label{gap}
\ee
It can be easily seen that,
when we approach the temperature $T^*$
where the mass $M(T)$ becomes zero, the stiffness
$J(T,\Lambda_T, M, N_c)$ also tends to zero.
Formula (\ref{Hei}) defines a generalized Heisenberg model
with a {\it temperature-dependent stiffness coefficient}.
The position of the phase disorder transition in a such
system should be determined self-consistently by
solving the system of  equations for $T_c$ and $M(T_c)$.
Apparently just as
in a superconductor
with a pseudogap,
the phase transition in  such a system is a competition
between the thermal depletion of the gap modulus (this
roughly corresponds to thermal pair breaking in a superconductor)
and the process of thermal excitations of
the directional  fluctuations in the
degenerate minimum of the effective potential. The
``BCS" limit corresponds to the situation where $T^*$ merges with
$T_c$ and it is easily seen that this scenario always holds true at
 $N_c \rightarrow \infty$. That is, at infinite $N$ the mean-field
theory is always accurate just as BCS theory works well in 
weak-coupling superconductors. In the framework
of this NLSM construction, at low $N_c$ the scenario
of the phase transition
depends on the  choice of $M(0), \Lambda_T $,
and ${\tilde \Lambda}_T$, which should be fixed from
phenomenological considerations.

%we introduced
%following to \cite{kb}
%a unit vector field
%$n_i
%\equiv (n_0,n_a)
%\equiv(\s,\pi_a)/ M $.

\section{Conclusion}
The precursor pairing fluctuations
is a general feature of any Fermi system
with composite bosons
and it is the dominant region of a phase
diagram of strong-coupling and low-carrier-density superconductors.
At the moment it is a subject of increasing interest in 
different branches of physics.
In the first part of this paper we briefly 
outlined the nonlinear sigma model 
approach to this phenomenon in superconductors 
in two and three dimensions.
In the second part we discussed  similar phomena
in the chiral Gross-Neveu and Nambu--Jona-Lasinio
models. This discussion should  have  relevance for
hot QCD and color superconductors.

We also note  that
in some sense similar phenomena are known
in a large variety of condensed matter systems,
in particular, besides superconductors
we can mention itinerant and local-momentum theories
of magnetism, exitonic condensate in
semiconductors, ferroelectrics and Josephson junction arrays \cite{knot}.

We would like to stress that the main purpose of this paper  is to
 summarize present discussions of precursor 
fluctuations in the Gross-Neveu and Nambu--Jona-Lasinio 
models.  We illustrated the discussion
with a few examples from superconductivity, outlining 
occurrence of similar phenomena in several arbitrarily chosen
models of superconductors with pseudogaps. Thus this paper
can not be regarded as a review of this phenomenon in superconductivity
which evolved recently to a very large branch of condensed matter 
physics. Thus our references to the papers on superconductivity 
are by definition incomplete, for more complete set of references reader 
may consult corresponding reviews on superconductivity [e.g. the review 
\cite{ranrev} ].

\comment{
Indication of possible importance of the pseudogap
concept in particle physics is the mentioned above existence of this
phenomenon in the chiral Gross-Neveu model at low $N$.
Even though these results can not be directly generalized to
NJL model, one can guess that in analogy
to 3D XY-model approach to strong-coupling
and low carrier density superconductivity, one
can set up a nonlinear 3D O(4)-sigma
model with temperature depended stiffness
coefficient as a toy model for QCD at finite temperatures
that would possess two characteristic temperatures
corresponding to discussed in this paper $T_c$ and $T^*$.
Speaking about the BCS-BEC crossover,
precritical fluctuations and the pseudogap phase,
it should be noted as well that
in some sense similar phenomena are known
in a large variety of condensed matter systems,
in particular, except for superconductors
we can mention itinerant and local-momentum theories
of magnetism, exitonic condensate in
semiconductors, ferroelectrics and Josephson junction arrays.}

%We also observed that due to nonuniversality of the
%critical stiffness the is {\it no} universal criterion
%for judgment of the nature
%of symmetry restoration in
%general systems as it was proposed in \cite{prl}
\begin{acknowledgments}
The author is grateful to Prof. T. Appelquist, Dr. V. Cheianov,
Prof. S. Hands, Prof. H. Kleinert,  Prof. A.J. Niemi  and Prof. 
 L.C.R. Wijewardhana 
for discussions and/or
 useful remarks, and to Prof. T. Hatsuda and Prof. D. Blaschke
for communicating  references.
\end{acknowledgments}


\begin{thebibliography}{999}
\bibitem{rem}
As we discussed in the first part of the paper, in a nonperturbative
NLSM approach to the  condensation from
pseudogap phase in strong-coupling superconductors one can reach
an excellent agreement with the perturbative results \cite{N,R}
if one considers an effective sigma model on the lattice
with the spacing given by $a \propto 1/n_{b}^{1/3}$,
where $n_b$ is  density of Cooper pairs \cite{sc}. Then
the model is related to the condensation of
hard-core composite bosons on a lattice.
\end{thebibliography}
\end{document}